# An Upper Bound to the Number of Gates on Single Qubit within One Error-Correction Period of Quantum Computation

Li Yang[*a] and Yufu Chen[b]

[a]State Key Laboratory of Information Security, Graduate University of Chinese Academy of Sciences, Beijing 100049, China;
[b]College of Mathematical Science, Graduate University of Chinese Academy of Sciences, Beijing 100049, China

**ABSTRACT**

Based on the amplitude behavior of quantum Rabi oscillation driven by a coherent field we show that there exists an upper bound to the number of logical operation performed on any single qubit within one error-correction period of a quantum computation. We introduce a parameter to depict the maximum of this number and estimate its decoherence limit. The analysis shows that a generally accepted error-rate threshold of quantum logic gates limits the parameter to so small a number that even a double of fault-tolerant Toffoli gates can hardly be implemented reliably within one error-correction period. This result suggests that the design of feasible fault-tolerant quantum circuits is still an arduous task.

**Keywords:** quantum computation, fault-tolerant quantum circuit, coherent field, decoherence limit, Rabi oscillation

## 1. INTRODUCTION

A quantum computer is a complicated quantum system which inevitably interacts with the environment and resulting in the failure of computation. It is obvious that any practical quantum computer has to incorporate some type of error correction into its operation [1]. The function fault-tolerance is particularly important to quantum computation since which implies that the computer can work effectively even when its elementary components are imperfect. Fault-tolerant quantum circuits are regarded as the most practical candidate for quantum computation [2, 3].

Coherent fields are usually used to drive Rabi rotation of two-level systems, or qubits, to realize operations of quantum gates [4-11]. Though each physical step of such gate operations contains only a half or mere a quarter Rabi period [4,5,8,9], the number of Rabi periods required in one error-correction period of fault-tolerant quantum computation is quite large [12,13], especially those computations by means of fault-tolerant Toffoli gate [2,11] and concatenated coding [12,13].

Our analysis is based on the fact that the quantum Rabi oscillation driven by a coherent field occurs under a certain Gaussian envelope [14], in order to give out an upper bound to the maximum number of gates on single qubit within one error-correction period theoretically [15]. It can be seen that this fact will lead the period number of Rabi oscillation to be limited since we require the oscillation is perfect enough to fit a given precision. If the fault-tolerant quantum circuit is so complicated that the number of perfect Rabi oscillations performed on any single qubit within one error-correction period is greater than the number of Rabi oscillations, the upper bound of error rate required by threshold theorem will be broken unavoidably, and the reliable quantum computation cannot be realized in principle, no matter what improvements in experiment with coherent fields carried out.

[*]E-mail address: lyang@m165.com

## 2. THE DEPTH OF LOGICAL OPERATION AND ITS DECOHERENCE LIMIT

The idea of fault-tolerant quantum computation is to compute directly on encoded quantum states, or logic qubits. A set of universal fault-tolerant quantum gates is, e.g., $\{\overline{H}, \overline{S}, \overline{CNOT}, \overline{Toff}\}$, here $\overline{H}$, $\overline{S}$, $\overline{CNOT}$ and $\overline{Toff}$ represent fault-tolerant constructions of Hadamard gate, phase gate, control-not gate and Toffoli gate respectively [2]. The construction of a quantum Toffoli gate with the $CNOT$ gate and single-qubit gates is sketched out in Figure 1, where $T$ represents $\dfrac{\pi}{8}$ gate and satisfies $T^2 = S$. This construction is essential because it is difficult to realize quantum Toffoli gate directly in a quantum system.

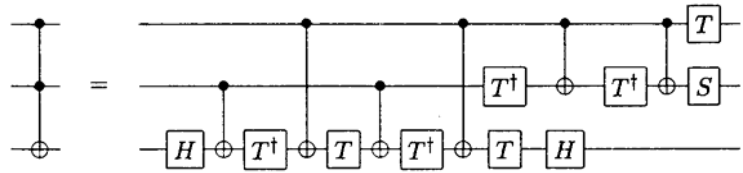

**Figure 1**. A construction of the quantum Toffoli gate based on the $CNOT$, T and S gate (After [11])

It can be seen that this construction includes 10 quantum gates on the target qubit. A construction of fault-tolerant Toffoli gate is much more complicated. The one pictured in Figure 2 is the original construction of Shor [2]. We can see that this typical construction of the fault-tolerant Toffoli gate contains 14 quantum gates totally on the third qubit.

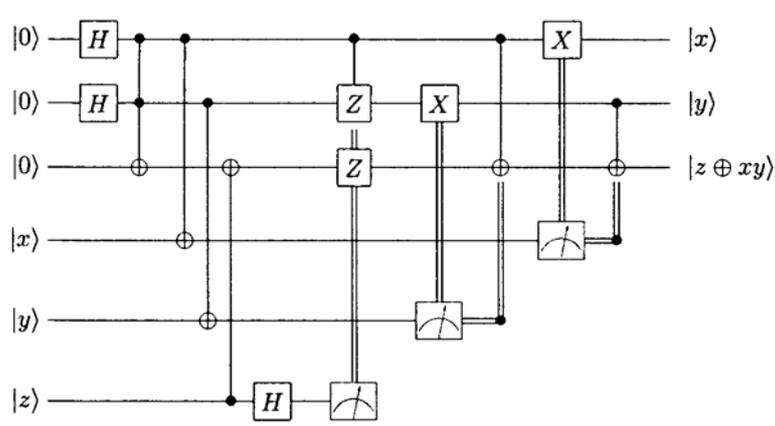

**Figure 2**. The original construction of quantum fault-tolerant Toffoli gate (After [11])

The threshold theorem of fault-tolerant quantum computation declares that an arbitrarily long quantum computation can be performed reliably if the error rate of each quantum gate is less than a critical value. One form of this theorem is as follows [11]:

**Threshold Theorem** A quantum circuit containing $p(n)$ gates may be simulated with probability of error at most $\varepsilon$ using

$$O\left(\text{poly}\left(\log\left(p(n)/\varepsilon\right)\right)p(n)\right) \tag{1}$$

Gates on hardware whose components fail with probability at most $p$, provided $p$ is below some constant *threshold*, $p < p_{th}$, and given reasonable assumptions about the noise in the underlying hardware.

The basis of the threshold theorem described above is the concatenated coding. It can be seen that the concatenated coding will lead the number of gates within one error-correction period to increase greatly [10]. We define the *depth of logical operation* of a quantum computation circuit as

$$\chi \equiv \max_{n \in I} x(n), \tag{2}$$

where $I$ represents the qubit set of the circuit, and $x(n)$ represents the number of logical operation on the nth qubit within one error-correction period. We can see that the parameter $\chi$ of the original construction of fault-tolerant Toffoli gate in Figure 2 is 14.

If the number of valid periods of quantum Rabi oscillation determined by the driving coherent field is smaller than $\chi$, the reliable fault-tolerant quantum computation declared by the threshold theorem will be not feasible in principle. We define this number the *decoherence limit* of parameter $\chi$. This limit is called a decoherence limit since it is set off by the decoherence of different Rabi oscillation amplitudes, and each of the amplitude arises from a different number state component of the driving field respectively. The decoherence limit under a reasonable driving coherent field and a generally accepted threshold will be estimated next section.

### 3. AN ESTIMATION OF THE DECOHERENCE LIMIT

Consider a two-level atom interacting with a coherent field, provided the atom is initially in the excited state. The probability for the atom to be found in the excited state at time $t$ is [14]

$$P(t) = \frac{1}{2}\left(1 + \sum_n \frac{e^{-\bar{n}}\bar{n}^n}{n!}\cos\left(2g\sqrt{n+1}\,t\right)\right), \tag{3}$$

where $\bar{n}$ is the average number of photons contained in a relevant coherent state pulse; $g \sim \dfrac{\varepsilon d}{\hbar}$, with $\varepsilon$ the field per photon, and $d$ the dipole moment of the two-level atom. Let $\tau \equiv gt$, and define

$$W(\tau) \equiv 2P\left(\frac{\tau}{g}\right) - 1 = \sum_n \frac{e^{-\bar{n}}\bar{n}^n}{n!}\cos\left(2\sqrt{n+1}\,\tau\right), \tag{4}$$

Numerical computation result of $W(\tau)$ is pictured in Figure 3. For $t < \sqrt{\bar{n}}/g$, a approximate formula of $P(t)$ is [14]

$$P(t) \approx \frac{1}{2}\left[1+\cos\left(2g(\bar{n}+1)^{1/2} t\right)\exp\left(-\frac{g^2 t^2 \bar{n}}{2(\bar{n}+1)}\right)\right]. \tag{5}$$

Thus, we have a approximate formula of $W(\tau)$ as

$$W(\tau) \approx \cos\left(2(\bar{n}+1)^{1/2}\tau\right)\exp\left(-\frac{\bar{n}}{2(\bar{n}+1)}\tau^2\right) \tag{6}$$

for $\tau < \sqrt{\bar{n}}$.

It can be seen by the approximate formula (4) that the envelope of $W(\tau)$ approaches $e^{-\frac{1}{2}\tau^2}$ as a limit while $\bar{n} \to \infty$. The angular frequency of average quantum Rabi oscillation is $\omega_\tau = 2\sqrt{\bar{n}+1}$, and the frequency is $\nu_\tau = \frac{\sqrt{\bar{n}+1}}{\pi}$. Thus we know the number of periods of Rabi oscillation given by approximate formula (6) in the interval $0 < \tau < 1$ with $\bar{n} = 10000$ is 31.8. The numerical result of formula (4) shows that this number is between 31.5 to 32 (See Figure 3), we can see that the expression of $\nu_\tau$ given above is a good approximation of the frequency of quantum Rabi oscillation.

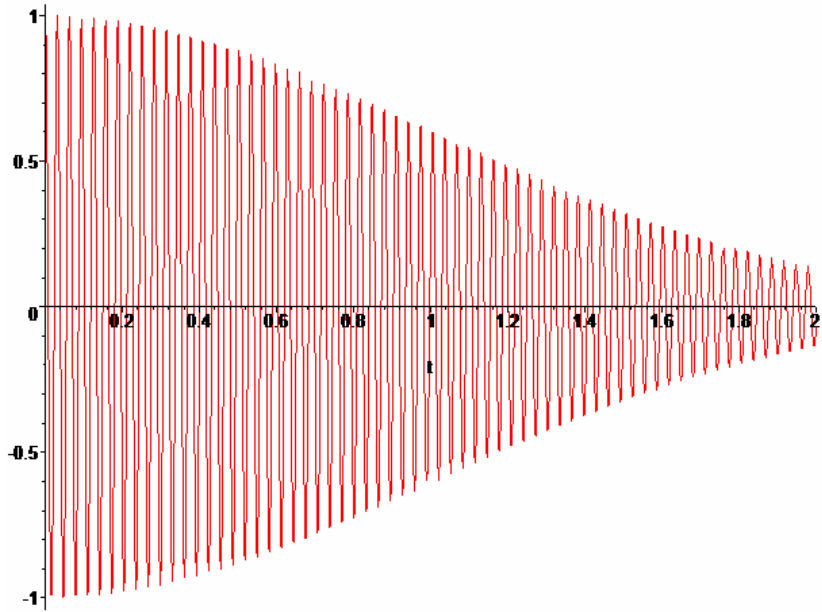

**Figure 3**. A numerical result of $W$ as a function of $\tau$, based on the definition in formula (4) with $\bar{n} = 10000$. This result shows visually that the quantum Rabi oscillations occur only under a Gaussian envelope, which is the start point of our analysis.

Suppose the precision required by threshold theorem is $p_{th} = 10^{-k}$. We require directly the amplitude of each valid period of Rabi oscillation must greater than $1 - p_{th}$ in order to get an evident result. Consider the following approximate inequality for the lower bound of the decoherence limit:

$$\sum_{i=1}^{N(\chi)} \frac{1}{2} \tau_i^2 < p_{th}, \tag{7}$$

where $N(\chi)$ is the lower bound of the number of valid periods of (average) quantum Rabi oscillation, or the lower bound of the decoherence limit of parameter $\chi$, and $\{\tau_i | i = 1, \cdots, N(\chi)\}$ the set of peak points of Rabi oscillation. Let $T$ represents the period of Rabi oscillation, we have $T = \frac{1}{\nu_\tau} = \frac{\pi}{\sqrt{\bar{n}+1}}$, then we have

$$2 p_{th} > T^2 \sum_{i=1}^{N(\chi)} i^2 = \frac{\pi^2}{6(\bar{n}+1)} N(\chi)(N(\chi)+1)(2N(\chi)+1) > \frac{\pi^2}{3(\bar{n}+1)} N^3(\chi),$$

thus we get

$$N(\chi) < \sqrt[3]{\frac{6(\bar{n}+1) p_{th}}{\pi^2}} = \sqrt[3]{\frac{6(\bar{n}+1) \times 10^{-k}}{\pi^2}}. \tag{8}$$

Now let us reckon the average number of photons contained in one coherent state pulse. Consider a case with wavelength of the coherent field $\lambda \sim 3 \times 10^{-7} \text{m}$, then the frequency is $\nu \sim 10^{15} \text{Hz}$, energy of each photon is $h\nu \sim 6 \times 10^{-19} \text{J}$. Provide quantum gates operate at a speed of $1 \text{ MHz}$, a reasonable width of control pulse is $\tau_p = 10^{-7} \text{s}$. Choose the peak power of the coherent pulse as $w_p \sim 10^{-3} \text{W}$, the energy of each coherent pulse is $\tau_p \times w_p = 10^{-10} \text{J}$. Thus we get

$$\bar{n} \approx \frac{\tau_p \times w_p}{h\nu} \sim \frac{5}{3} \times 10^8. \tag{9}$$

Let $p_{th} = 10^{-4}$ [16], then we have

$$N(\chi) < \sqrt[3]{\frac{6 \times \frac{5}{3} \times 10^8}{\pi^2} \times 10^{-4}} = 10 \times \sqrt[3]{10^2/\pi^2} \approx 21.6 \ . \tag{10}$$

This lower bound is too small to executing the computation of one error-correction period of fault-tolerant quantum circuits.

Consider the case in an experiment [5]: $\nu \sim 9 \times 10^{14}\,\text{Hz}$, $\tau_p \sim 10^{-4}\,s$, $w_p \sim 10^{-3}\,\text{W}$, then $h\nu \sim 6 \times 10^{-19}\,\text{J}$, $\tau_p \times w_p = 10^{-7}\,\text{J}$, and $\bar{n} \approx \frac{\tau_p \times w_p}{h\nu} \sim \frac{1}{6} \times 10^{12}$. We get

$$N(\chi) < \sqrt[3]{\frac{6 \times \frac{1}{6} \times 10^{12}}{\pi^2} \times 10^{-4}} \approx 216.$$

This result is much better than that in (10), but the relevant switching speed of the CNOT gate is 20 kHz [5].

An alternative construction of quantum fault-tolerant Toffoli gate [16] is pictured in Figure 4. We can see that the parameter $\chi$ of this construction is 12. It is possible that the parameter $\chi$ of any construction of the quantum fault-tolerant Toffoli gate is greater than 10. Perhaps it is suitable to realize the quantum Toffoli gate directly with the scheme suggested in [4], though it is difficult in experiment.

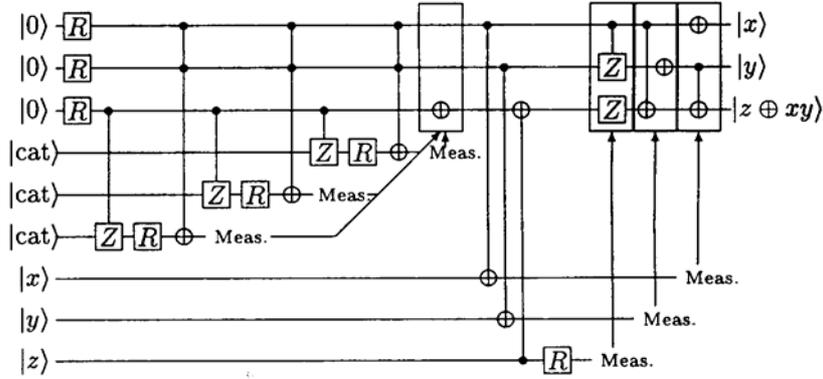

**Figure 4.** Another construction of quantum fault-tolerant Toffoli gate (After [1])

E. Knill has presented a $C_4/C_6$ architecture [13] of fault-tolerant quantum computation circuits with a threshold about 1%. By means of the formula (8) and parameters given above, provided the switching speed of the CNOT gate is 1MHz, we get $N(\chi) < 100$. We have to use at lest $10^9$ s, or 30 years, to implement the $2 \times 10^{17}$ physical $CNOT$ s of a single realistic computation task [13] even taking into account of the parallel performance. Moreover, we do not know whether $\chi \sim 100$ is sufficient for implementing all the computations of one error-correction period of a fault-tolerant quantum circuit based on the $C_4 / C_6$ architecture.

One way left to overcome the difficulty caused by small $N(\chi)$ is increasing the peak power of relevant coherent state pulses, which will unavoidably raise the computation noise and then limited by a given threshold $p_{th}$. For example, in the physical realization scheme with cold trapped ions, sufficiently low intensities of laser pulses is necessary to control the interactions between the ions through the center-of-mass motion [4]. We can see that the scale of one error-correction period of a quantum computation driven by coherent fields is limited in principle, regardless of improvements in experiment. It is worth to mention that the time of executing the whole computation is inverse proportion to the number of periods divided.

## 4. CONCLUSIONS

We have introduced a parameter $\chi$ called the depth of logical operation to depict a fault-tolerant quantum computation circuit. Based on the amplitude behavior of quantum Rabi oscillation driven by a coherent field we get the decoherence limit of the depth $\chi$. A necessary condition of executing a quantum circuit reliably is that the parameter $\chi$ does not exceed its decoherence limit.

The decoherence limit of logical operation depth to a quantum computer with a gate operation rate $1\,\mathrm{MHz}$ and an error rate less than $10^{-4}$ is only about 21. This result implies that we cannot execute a double of quantum fault-tolerant Toffoli gates within one error-correction period, no matter what improvements in experimental techniques carried out. In the case of $C_4 / C_6$ architecture with $p_{th} \sim 1\%$, we show that the decoherence limit is about 100, provided the switching speed of the quantum gate is $1\,\mathrm{MHz}$. We have not checked whether $N(\chi) < 100$ is enough for executing the computation within one error-correction period of concatenated $C_4 / C_6$ architecture yet. Perhaps the design of fault-tolerant circuits, the physical realization of quantum gates and the optimization of quantum algorithms should be considered comprehensively to realize both reliable and practical fault-tolerant quantum computation in future.

## ACKNOWLEDGEMENT

We would like to thank G. L. Long for useful discussions. This work was supported by the National Natural Science Foundation of China under Grant No.60573051, and the Fundamental Research Program of China under Grant

No. 2001CB309300.